# The satellite archaeological survey of Egypt

Amelia Carolina Sparavigna
Dipartimento di Fisica,
Politecnico di Torino, Torino, Italy

A recent announcement of some pyramids, buried under the sand of Egypt and discovered by means of infrared remote sensing, renewed the interest on the archaeological surveys aided by satellites. Here we propose the use of images, obtained from those of Google Maps after some processing to enhance their details, to locate archaeological remains in Egypt.

A recent announcement from BBC of 17 new pyramids discovered in Egypt arouses the interest on the archaeology aided by satellites imagery [1]. These pyramids, as many other ancient remains in Egypt, are under the sand of the desert. They were discovered by means of a remote sensing with infrared sensors. In fact, the archaeological surveys, usually performed by means of airplanes, are necessary to observing the sites from above and gain a better view of the landforms. In some cases, the survey of a region ends with the discovery of new archaeological sites or with the precise location of an ancient lost town [2].
Satellites give different opportunities, according to their sensorial equipment. BBC announced that Sarah Parcak, of the University of Alabama, used some data from NASA infrared equipped satellites to survey the Egypt. Waiting for a more detailed report on her researches and on the methods the team used, we can just tell that the infrared inspection is based on collecting the radiances in various wavelength bands, in the infrared range of the electromagnetic spectrum. The resulting profiles depend on the methods used to obtain the surface data from radiances. To have a good detection, the surface must be free from clouds.
The Egypt's Minister of State for Antiquities Affairs, Zahi Hawass, seems to be quite interested to the new technologies, but, as he told Ahram Online, the satellite infrared images are only able to locate the remains beneath the sand [3]. It is then necessary, according to Hawass, to identify them with archaeological researches on the spot. From the news on the Web it is not clear how many sites have been analyzed by the team of the University of Alabama. It seems that the amount of data collected by the researchers is huge.
Besides the analysis with infrared imagery, let us consider that there are other remote sensing techniques that can be useful in archaeology: among them we have the LIDAR system, which is, as we discussed in [4], able to see under the canopy of the forests, and the SIR-C/X-SAR imaging radar system, which has waves that can penetrate the clouds, and, under certain conditions, vegetation, ice and dry sand [5]. Of course, these facilities are not freely available and needs financial supports.
We could then ask ourselves if a free satellite service, such as Google Maps, can help in some archaeological researches in Egypt. It is my opinion that the answer is positive. In studying the Merowe Dam and the paleochannels of the Nile we could compare the images from SIR-C/X-SAR imaging radar system, with those of the Google Maps [6]. After a suitable image processing with some freely downloadable programs (GIMP, IRIS, AstroFracTool,[7]), the Google Maps revealed astonishing details of the network of old buried channels of Nile in the Nubian region. The same for the "raised fields" near the Titicaca Lake in Peru: the processing of the images clearly displayed the network of these ancient earthworks and canals [8]. Many of these structures are probably buries under some sediments of the lake.
Let us then try to apply the image processing to the Google Maps of those areas in Egypt, where according to the press, the infrared satellite imagery is giving good results. We see that one of these investigated areas is that of Tanis, a town of the ancient Egypt. In Fig.1, it is shown what we can have after processing the image from Google Maps. The upper part of the figure is obtained using the GIMP image- processing program, to adjust brightness and contrast. The lower part is gained after a processing with the wavelet filtering of Iris. These images seem to contain quite clear information on the buried town too.

Another example is the site where there are buried pyramids, according to the press [9,10]. The site is at Saqqara: Figure 2 shows the area as can be seen after a processing of Google Maps. The reader is invited to compare these images with those published on the Web, copyrighted BBC. I guess that after comparison, the reader can draw some positive conclusions about Google Maps and its use for an archaeological survey of Egypt. I am proposing another example of the use of image processing in Fig.3. This is the Great Temple at Amarna, buried under the sand (more images at [11]).

As Zahi Hawass is telling, it is necessary to understand whether some "anomalies" revealed by the satellite remote sensing are archaeological remains or not. This means that archaeology can only receive benefits for geophysics researches and the related use of remote sensing.

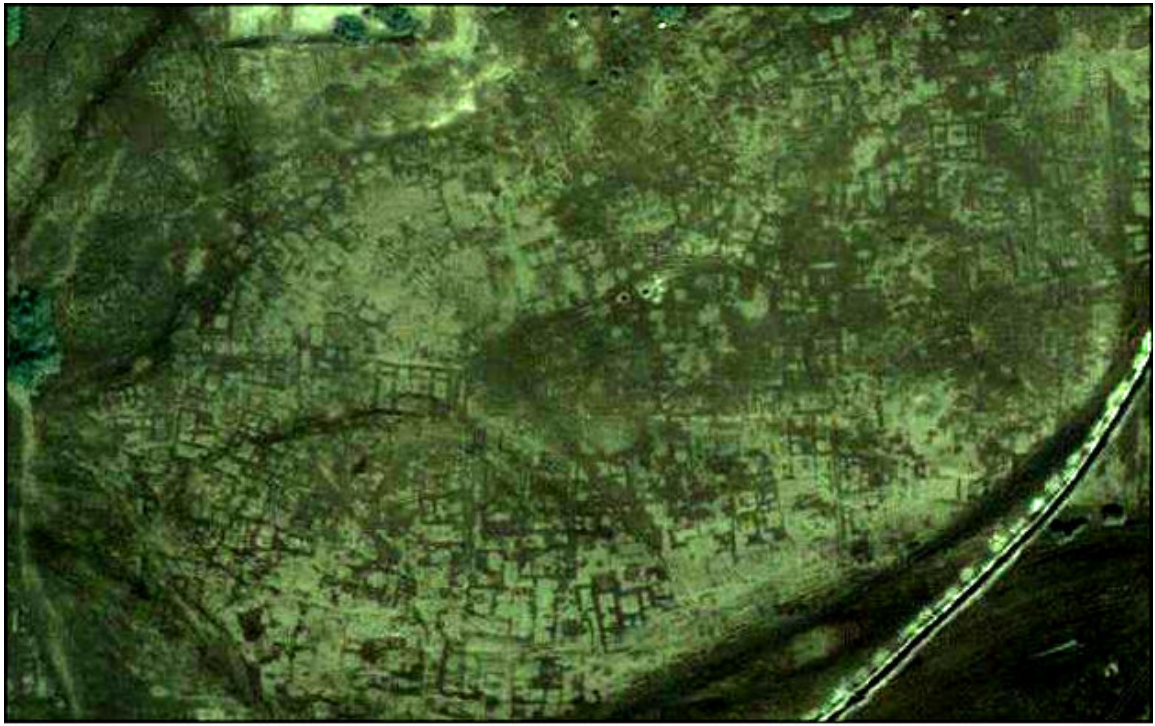
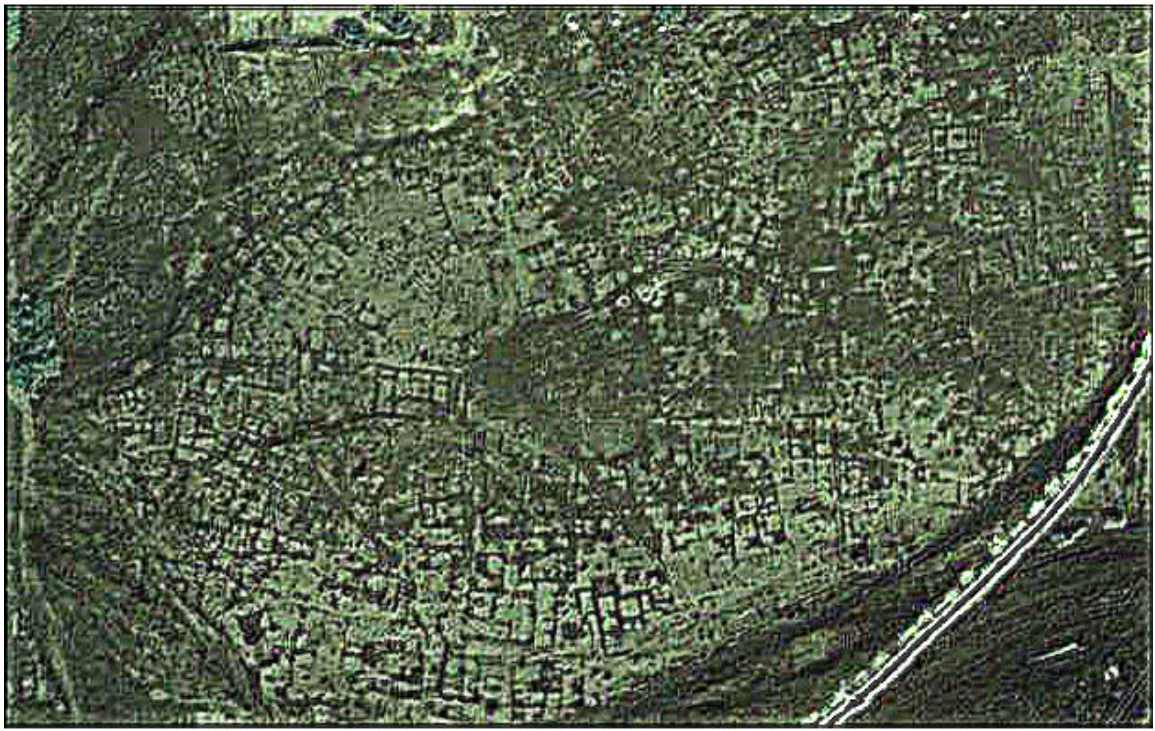

Fig.1: Tanis as can be observed after processing an image from Google Maps. In the upper part, brightness and contrast had been adjusted with GIMP. The lower image was obtained with a wavelet filtering with Iris. It seems that the image is giving quite good details too.

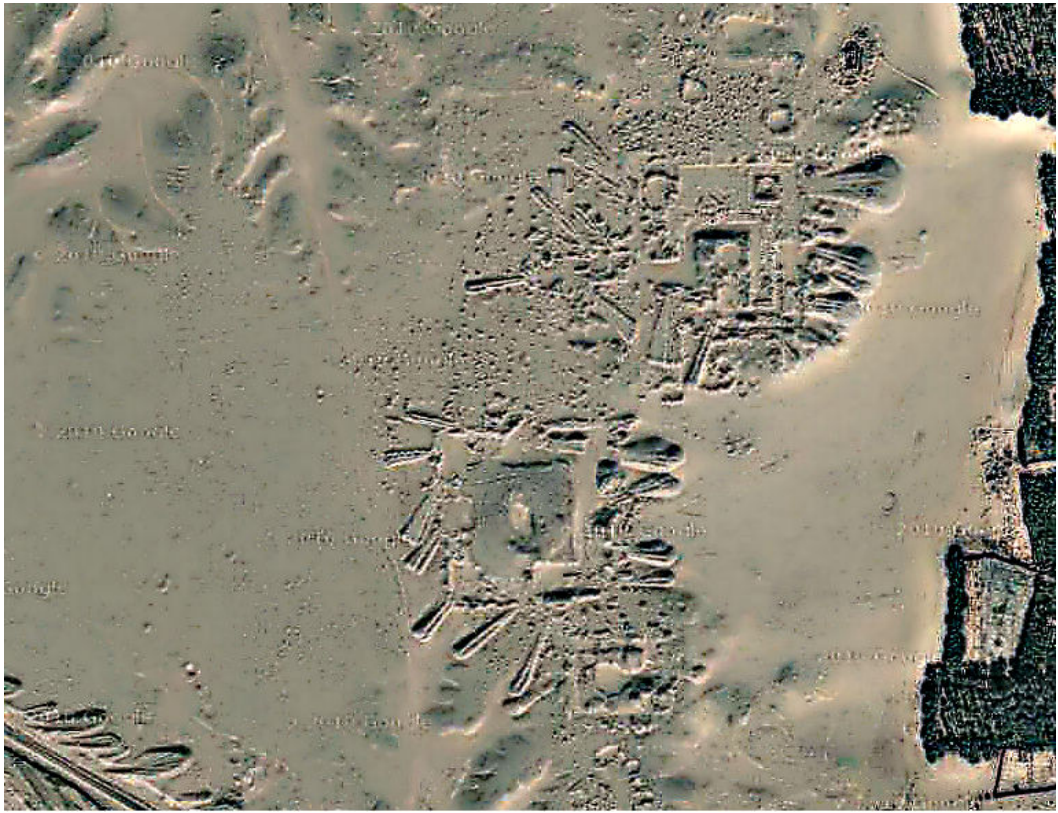

Fig.2. This is the Saqqara area where there is a buried pyramid. The image has been obtained after processing a Google Maps image. According to Refs.[9], in this area there one of the buried pyramid announced by BBC [1].

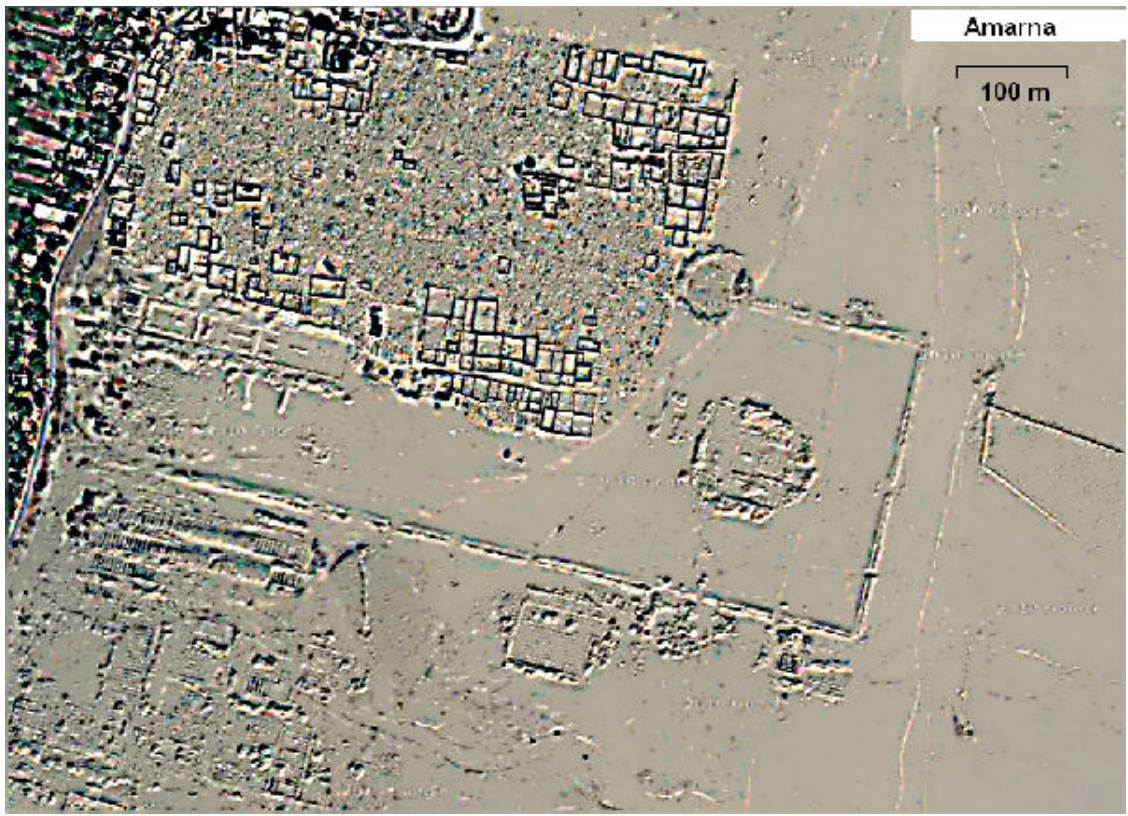

Fig.3: The Great Temple in Amarna, as can be seen after processing a Google Maps image.